# A Quasi-Classical Mapping Approach to Vibrationally Coupled Electron Transport in Molecular Junctions

Bin Li,[a] Eli Y. Wilner,[b] Michael Thoss,[c] Eran Rabani,[d] and William H. Miller[a]

*(a) Department of Chemistry and Kenneth S. Pitzer Center for Theoretical Chemistry, University of California, and Chemical Sciences Division, Lawrence Berkeley National Laboratory, Berkeley, California 94720, USA*

*(b) School of Physics and Astronomy, The Sackler Faculty of Exact Sciences, Tel Aviv University, Tel Aviv 69978, Israel*

*(c) Institute for Theoretical Physics and Interdisciplinary Center for Molecular Materials, Friedrich-Alexander-Universität Erlangen-Nürnberg, Staudtstr. 7/B2, 91058 Erlangen, Germany*

*(d) School of Chemistry, The Sackler Faculty of Exact Sciences, Tel Aviv University, Tel Aviv 69978, Israel*

Abstract: We develop a classical mapping approach suitable to describe vibrationally coupled charge transport in molecular junctions based on the Cartesian mapping for many-electron systems [*J. Chem. Phys*. **137**, 154107 (2012)]. To properly describe vibrational quantum effects in the transport characteristics, we introduce a simple transformation rewriting the Hamiltonian in terms of occupation numbers and use a binning function to facilitate quantization. The approach provides accurate results for the nonequilibrium Holstein model for a range of bias voltages, vibrational frequencies and temperatures. It also captures the hallmarks of vibrational quantum effects apparent in step-like structure in the current-voltage characteristics at low temperatures as well as the phenomenon of Franck-Condon blockade.

## I. INTRODUCTION

The study and understanding of quantum transport processes in molecular nanostructures has been of great interest recently. Among the variety of architectures and processes considered, charge transport in single molecule junctions, i.e. molecules chemically bound to metal or semiconductor electrodes, has received particular attention.[1-4] These systems combine the possibility to study fundamental aspects of nonequilibrium many-body quantum physics at the nanoscale with the perspective for technological applications in nanoelectronic devices. Recent experimental studies of transport in molecular junctions have revealed a wealth of interesting transport phenomena such as Coulomb blockade, Kondo effects, negative differential resistance, transistor- or diode-like behavior, as well as switching and hysteresis.[5-13] An important aspect that distinguishes molecular conductors from mesoscopic semiconductor devices is the influence of the nuclear degrees of freedom (DoF) of the molecular bridge on transport properties.[14–16] Due to the small size of molecules, the charging of the molecular bridge is often accompanied by significant changes of the nuclear geometry, indicating strong electronic-vibrational (vibronic) coupling. This coupling manifests itself in vibronic structures in the transport characteristics and may result in a multitude of nonequilibrium phenomena such as current induced local heating and cooling, multistability, switching and hysteresis, as well as decoherence.[6,14-35]

A variety of theoretical methods have been developed and applied to treat vibrationally coupled charge transport in molecular junctions. Examples of approximate methods are scattering theory,[36-42] nonequilibrium Greens function (NEGF) approaches,[17,43-54] and master equation methods.[44,55-63] In addition, a variety of numerically exact schemes have employed, including numerical path-integral approaches,[64-66] the multilayer multiconfiguration time-dependent Hartree (ML-MCTDH) method,[67,68] the scattering state numerical renormalization group approach,[69] and a combination of reduced density matrix techniques and impurity solvers.[32,70]

All these methods employ a quantum mechanical treatment of both the electronic and nuclear DoF. An alternative strategy is to use classical concepts, which typically scale much more favorably with the dimensionality of the problem, i.e. the number of nuclear DoF, than fully quantum mechanical methods, and also allow a straightforward application to systems with anharmonic potential energy surfaces, which is a challenge, e.g., for NEGF theory and path-integral methods. To apply classical concepts to systems with electronic and nuclear DoF, it is advantageous to treat the overall system on an equal dynamical (i.e. classical) footing,[71-74] which requires a classical model for the electronic DoF. This can be obtained by mapping the discrete electronic DoF (i.e. electronic states) onto continuous DoF. Subsequently, any classical or semiclassical treatment can be employed for the overall system, i.e. the electronic and nuclear DoF. In the related field of electronically nonadiabatic molecular dynamics, such a mapping is achieved, e.g., by using the Meyer-Miller-Stock-Thoss (MMST) method, which represents electronic states by harmonic oscillators.[72,75-80] Recently, similar mapping strategies have been proposed for the many-electron problem inherent in the transport scenario. Swenson et al.[81] have adopted a mapping approach based on the early work of Miller and White,[82] constructing a classical Hamiltonian corresponding to the general second-quantized Hamiltonian operator for a many-electron system in which all the creation and annihilation operators for the spin orbitals were substituted by a set of classical action angle variables. This scheme was employed to calculate the current of the resonant level model with and without electron-vibrational coupling,[81,83] ignoring in both cases electron-electron interactions. This mapping approach provides semiquantitative results for a range of system parameters, i.e., different source-drain voltages, temperatures, and electron-vibrational couplings. However, it falls short in providing accurate and even qualitative results in several different limits. For example, it fails to capture the effects of a gate voltage in the noninteracting case, as well as vibronic Franck-Condon structures in the transport characteristics for the interacting case, which are ubiquitous in systems with small molecule-lead coupling and small to moderate vibrational relaxation. Moreover, it does not fully capture the Coulomb blockade when electron-electron interactions are introduced.

To overcome these shortcomings, recently a new approach was proposed for the many electron second quantized Hamiltonian.[84] This approach employs the representation of products of fermionic creation and annihilation operators by quaternions and the formal relation between quaternions and vector cross products. Specifically, the single-electron coupling operators are mapped to the cross product of coordinate and their conjugate momentum vectors, both represented in Cartesian coordinates. This kind of mapping naturally preserves all anti-commutation relationships of the fermionic creation and annihilation operators in the classical Hamiltonian. Within a semiclassical initial value representation (SC-IVR) treatment, it was shown to be exact for purely electronic problems (i.e. fixed nuclei) if only single-electron terms are included and to provide accurate results when two-electron terms are included.[84] For nonequilibrium dynamics in a resonant level model without electron-vibrational and electron-electron interaction, the quasi-classical implementation of the approach provides accurate results even when a finite gate voltage is applied, correcting the flaws of the previous action-angle mapping.[85] Moreover, it qualitatively captures the Coulomb blockade observed for large electron-electron interactions as shown for the Anderson impurity model.[85]

In the present paper, we apply this new mapping approach within a quasi-classical implementation to vibrationally coupled electron transport employing the Holstein model, which involves an electronic level at the molecular bridge coupled to a vibrational mode and to fermionic leads representing the electrodes. Furthermore, we devise an optimized representation of the Hamiltonian and a binning method to capture the vibronic Franck-Condon structures in the transport characteristics, which are due to quantization of the vibrational DoF in the molecular resonance state and which are missed in traditional quasi-classical treatments.

This paper is arranged as follows: Section II introduces the model, the classical mapping method used to represent the second quantized Hamiltonian, as well as the quasi-classical implementation. Furthermore, a special representation of the Hamiltonian is devised, which is optimized to describe vibronic Franck-Condon structures within a quasi-classical approach. Results of the quasi-classical approach for different parameter regimes are presented in Section III. Comparison with numerically exact results from real-time path integral Monte Carlo (RT-PIMC) simulations illustrates the performance of the approach. Section IV concludes with a summary.

## II. THEORY

### A. Hamiltonian and model parameters

To study vibrationally coupled electron transport in a nanostructure such as a molecular junction or a quantum dot we use the Holstein model, which involves an electronic level at the molecular bridge (in the following referred to as the 'dot') coupled to a vibrational mode and to fermionic leads representing the electrodes. The second quantized Hamiltonian for the Holstein model can be expressed in terms of a sum over the dot Hamiltonian $H_D$, leads Hamiltonian $H_{leads}$ and the couplings between the dot and the leads V:

$$H = H_D + H_{lead} + V, \quad (1)$$

where

$$H_D = \left(\varepsilon_D + \frac{\gamma^2}{\omega}\right)\hat{d}^\dagger\hat{d} + \omega\hat{b}^\dagger\hat{b} + \gamma\hat{d}^\dagger\hat{d}(\hat{b}^\dagger + \hat{b}), \quad (2)$$

$$H_{lead} = \sum_{k \in L,R} \varepsilon_k \hat{c}_k^\dagger \hat{c}_k, \quad (3)$$

and

$$V = \sum_{k \in L,R} t_k (\hat{d}^\dagger \hat{c}_k + \hat{c}_k^\dagger \hat{d}). \quad (4)$$

Thereby, $\hat{d}^\dagger/\hat{d}$ and $\hat{c}_k^\dagger/\hat{c}_k$ are fermionic creation/annihilation operators for the dot and the leads and $\hat{b}^\dagger/\hat{b}$ are bosonic creation/annihilation operators of the vibrational mode. The dot-lead interaction strength $t_k$ is determined by the spectral density

$$J_{L/R}(\varepsilon) = 2\pi \sum_{k \in L,R} |t_k|^2 \delta(\varepsilon - \varepsilon_k) \quad (5)$$

In the present paper, the following functional form is used for the spectral density

$$J_{L/R}(\varepsilon) = \frac{\Gamma_{L/R}}{[1+e^{A(\varepsilon-B)}][1+e^{-A(\varepsilon+B)}]} \tag{6}$$

which involves a cut-off outside the region of interest. In the numerical calculations, the spectral density is discretized on an equidistant grid, which gives the molecule lead coupling parameters

$$t_k = \sqrt{\frac{J(\varepsilon_k)\Delta\varepsilon}{2\pi}}. \tag{7}$$

Specifically, the left and right leads were discretized with 1000 states, resulting in a discretization interval of $\Delta\varepsilon = 0.03\Gamma$. In all calculations, a symmetric coupling to left and right leads is considered, i.e. $\Gamma_L = \Gamma_R = \frac{1}{2}\Gamma$, with a cut-off width of $A = 5\,\Gamma$ and a cut-off range $B = 10\Gamma$.

## B. Cartesian mapping

The Cartesian mapping has been introduced in Ref. 84 and also described in detail for the Anderson impurity model in Ref. 85. Here, we briefly describe the mapping approach for the Holstein model.

The fermionic creation and annihilation operators, isomorphic to quaternions,[84] can be mapped to the z-component of a cross product of a coordinate $\boldsymbol{r}$ and a conjugate momentum $\boldsymbol{p}$. For the populations of the dot and leads, the mapping is given by:

$$\hat{d}^\dagger \hat{d} \to \tfrac{1}{2} + xp_y - yp_x, \tag{8}$$

and

$$\hat{c}_k^\dagger \hat{c}_k \to \tfrac{1}{2} + x_k p_{yk} - y_k p_{xk}, \tag{9}$$

respectively. The dot-lead couplings are mapped according to:

$$\hat{c}_k^\dagger \hat{d} + \hat{d}^\dagger \hat{c}_k \to x_k p_y - y_k p_x + x p_{yk} - y p_{xk}. \tag{10}$$

The Holstein model includes also bosonic operators, which are represented in the usual way by dimensionless Cartesian coordinates and momenta, i.e. $\hat{b}^\dagger + \hat{b} = \sqrt{2}x_b$ and $\hat{b}^\dagger \hat{b} = \frac{1}{2}(x_b^2 + p_b^2 - 1)$. Using these definitions, the dot Hamiltonian is given by:

$$H_D = \left(\varepsilon_D + \tfrac{\gamma^2}{\omega}\right)(xp_y - yp_x) + \tfrac{\omega}{2}(x_b^2 + p_b^2 - 1) + \sqrt{2}\gamma(xp_y - yp_x)x_b. \tag{11}$$

Similarly, the lead and coupling terms are given by:

$$H_{lead} = \sum_{k \in L,R} \varepsilon_k (x_k p_{yk} - y_k p_{xk}) \tag{12}$$

and

$$V = \sum_k t_k (xp_{yk} - yp_{xk} + x_k p_y - y_k p_x), \tag{13}$$

respectively.

In what follows, we will mainly be interested in the calculation of the time dependent current and its steady-state value. Using the Cartesian mapping, the current from the left/right lead to the dot is given by:

$$I_{L(R)} = -e\frac{d}{dt}\sum_{k \in L(R)} t_k(xp_{yk} - yp_{xk} + x_k p_y - y_k p_x) = \sum_{k \in L(R)} t_k(yp_{yk} + xp_{xk} - x_k p_x - y_k p_y), \tag{14}$$

and the total current is defined as $I_{tot} = \frac{1}{2}(I_L - I_R)$.

## C. Quasi-classical approximation

The above mapped Hamiltonian is a starting point for semiclassical and quasi-classical approximations. Here, we limit the treatment to a quasi-classical approach, where all DoF are propagated classically using Hamilton's equation of motion. Within this approach, it is straightforward to show that the total energy ($H_D + H_{lead} + V$) as well as the total number of electron $(x_d p_{yd} - y_d p_{xd} + \sum_{k \in L,R}(x_k p_{yk} - y_k p_{xk}))$ are conserved, as they should be.

We adopt a quasi-classical procedure to determine the initial conditions of the coordinates and their conjugate momenta, which mimics the correct quantum partition function of the noninteracting system.[81,83,85] We enforce quantum statistics on the initial conditions for each DoF by setting the initial occupation to either 0 or 1 such that its expectation value, averaged over the set of initial conditions, satisfies the Fermi-Dirac distribution:

$$n_k = \begin{cases} 0 & \xi_k > [1 + e^{\beta(\epsilon_k - \mu_{L/R})}]^{-1} \\ 1 & \xi_k \leq [1 + e^{\beta(\epsilon_k - \mu_{L/R})}]^{-1}, \end{cases} \quad (15)$$

where $\xi_k$ is a random number in the interval [0,1], $\beta$ is the inverse of temperature and $\mu_{L/R}$ is the left/right chemical potential, such that $\mu_L = -\mu_R = \frac{1}{2}V_{s-d}$ ($V_{s-d}$ is the source-drain voltage). The phase space variables are then sampled according to:[85]

$$\begin{aligned} x_k &= r_k \cos\theta_k, & p_{xk} &= p_{rk}\cos\theta_k - n_k \frac{\sin\theta_k}{r_k}, \\ y_k &= r_k \sin\theta_k, & p_{yk} &= p_{rk}\sin\theta_k + n_k \frac{\cos\theta_k}{r_k}. \end{aligned} \quad (16)$$

In the above, $n_k$ is defined by Eq. (15) and $\theta_k$ is a random variable in the interval $[0,2\pi]$. We find that the results converge rapidly with $r_k = 1$ and $p_{rk} = 0$, which is the choice used for the results reported below.

For the vibrational variables, we use the Wigner transform of the Boltzmann operator and sample $x_b$ and $p_b$ from:

$$\rho_w(x_b, p_b) = \frac{1}{Q}\exp\left\{-\tanh\left(\frac{\hbar\beta\omega}{2}\right)(x_b^2 + p_b^2)\right\}, \quad (17)$$

where $Q = \cosh\left(\frac{\hbar\beta\omega}{2}\right)$.

### D. Special treatment of the electron-vibrational couplings

As is well known, for a Holstein model with a symmetric drop of the source-drain bias on the junction, the steady-state current shows a step-wise increase when $V_{s-d}$ approaches a threshold of $2(\varepsilon_D + j\omega)$, where $j$ is an integer. These steps mark the onset of transport channels involving vibronic transitions from the vibrational states of the uncharged dot to those of the charged dot. The height of the steps are, to a first approximation, described by the corresponding Franck-Condon factors.[3,14,15,86,87] The resulting suppression of the current for small bias voltages, which is particularly pronounced for large vibronic coupling, has been termed Franck-Condon blockade,[86,87] in analogy to the phenomenon of Coulomb blockade. These vibronic Franck-Condon features are a manifestation of the quantization of the vibrational DoF in the charged molecular state and cannot be captured by a direct classical simulation of the Hamiltonian given by Eqs. (11)-(13) with the quasi-classical choice of initial conditions, where the number of vibrational quanta assumes a continuous value. Fig. 1 demonstrates this failure for a specific choice of the model parameters. The classical simulations are compared to converged results using the RT-PIMC approach of Ref. 66.

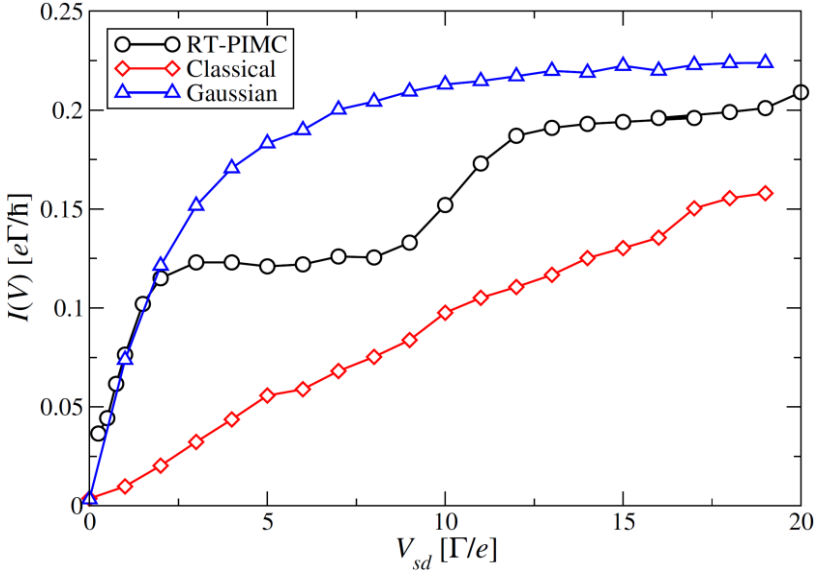

*Figure 1: Steady-state current for the original classical mapping, compared with the exact quantum results (RT-PIMC). In addition, results employing a Gaussian binning with $\delta n = \frac{1}{100}$ are shown (see the main text for details). Model parameters are $\omega = 5, \gamma = 4, \beta = 5$ (in units of $\Gamma$), and $\varepsilon_D = 0$.*

Even when we use a simple correction to the dot occupations which provided accurate results for the Coulomb blockade in the Anderson impurity model,[85] the approach fails to reproduce the vibronic Frank-Condon features. Using the mapping relation $\hat{d}^\dagger \hat{d} = \lim_{\delta n \to 0} \exp\left\{-\frac{(\hat{d}^\dagger \hat{d} - 1)^2}{\delta n}\right\}$ for the electron-vibrational interaction term in the Hamiltonian, Eq. (2), indeed improves the low-bias behavior (see blue curve in Fig. 1, but still fails completely to recover the step-like structure in the I-V plot.

To capture the vibronic Frank-Condon features, we rewrite the dot Hamiltonian given by Eq. (2) as follows:

$$H_D = \varepsilon_D \hat{d}^\dagger \hat{d} + \omega \hat{d}^\dagger \hat{d} \left(\hat{b}^\dagger + \frac{\gamma}{\omega}\right)\left(\hat{b} + \frac{\gamma}{\omega}\right) + \omega(1 - \hat{d}^\dagger \hat{d})\hat{b}^\dagger \hat{b} \tag{18}$$

and introduce a new set of shifted bosonic creation and annihilation operators:

$$\tilde{\hat{b}}^\dagger = \hat{b}^\dagger + \frac{\gamma}{\omega},$$
$$\tilde{\hat{b}} = \hat{b} + \frac{\gamma}{\omega}. \tag{19}$$

It is clear from Eq. (18) that the dot Hamiltonian describes two diabatic harmonic potential energy surfaces, one centered at the origin (corresponding to the uncharged state, i.e. $\hat{d}^\dagger \hat{d} = 0$) and the other at $-\frac{\gamma}{\omega}$ (corresponding to the charged state, $\hat{d}^\dagger \hat{d} = 1$). Therefore, the original operators, $\hat{b}^\dagger$, $\hat{b}$, are natural to describe the motion on the diabatic surface corresponding to $\hat{d}^\dagger \hat{d} = 0$ while the new set defined by Eq. (19) is natural for the shifted diabatic surface ($\hat{d}^\dagger \hat{d} = 1$). The vibrational number operator,

$$n_b = \hat{b}^\dagger \hat{b} = \frac{1}{2}(x_b^2 + p_b^2 - 1) \tag{20}$$

can also be defined for the charged state and reads

$$\tilde{n}_b = \tilde{\hat{b}}^\dagger \tilde{\hat{b}} = \left[\frac{1}{2}(x_b^2 + p_b^2 - 1) + \sqrt{2}\frac{\gamma}{\omega}x_b + \left(\frac{\gamma}{\omega}\right)^2\right]. \tag{21}$$

Using these definitions, the dot Hamiltonian in Eq. (18) can be written in terms of both shifted and unshifted number operators:

$$H_D = \varepsilon_D \hat{d}^\dagger \hat{d} + \omega \hat{d}^\dagger \hat{d} \tilde{\hat{b}}^\dagger \tilde{\hat{b}} + \omega(1 - \hat{d}^\dagger \hat{d})\hat{b}^\dagger \hat{b}. \tag{22}$$

This form is now suitable for introducing a binning function for the quantization of the vibrations, since it only depends on the vibrational occupation operators rather than the individual creation and annihilation operators. For the Anderson impurity model we used a narrow Gaussian to describe the two-body term in the Hamiltonian,[85] as shown in Fig. 1. This form worked well for the electronic DoF, but is not suitable for the vibrations, since their occupation can exceed the value of unity. Thus, we propose the following multistage function to "quantize" the vibrations:

$$f(n_b) = \sum_{v=1} \theta\left(n_b - v + \frac{1}{2}\right) \qquad (23)$$

where $\theta(x)$ is the Heaviside function approximated in the numerical calculations by $\theta(x) = (1 + e^{-\lambda x})^{-1}$ and $\lambda$ is a parameter.

This multistage function is adopted for both $n_b$ and $\tilde{n}_b$ in Eq. (22), and the resulting mapped dot Hamiltonian is given by

$$H_D = \varepsilon_D(xp_y - yp_x) + \omega f(n_b) + \omega(xp_y - yp_x)(f(\tilde{n}_b) - f(n_b)), \qquad (24)$$

with $n_b$ and $\tilde{n}_b$ given by Eqs. (20) and (21) respectively. The above form of the dot Hamiltonian is the one used for the applications reported below.

## III. RESULTS

To test the accuracy of the proposed mapping, we have carried out simulations for various model parameters ($\omega$, $\beta$ and $\mu_{L,R}$) and compared the results to numerically exact quantum calculations based on the RT-PIMC approach.[66] Approximately 320,000 trajectories were required to converge the values of the time-dependent and steady-state current to a desired accuracy. The dot population was set to $n_d = 0$ (unoccupied dot) initially. The initial state of the electrons in the leads and the vibrational mode were sampled according to Eqs. (15)-(17), respectively.

In Fig. 2 we show results for the time-dependent currents for two different values of the vibrational frequency and for two different values of the source-drain bias voltage. The RT-PIMC results are hard to converge for times beyond $t > 5/\Gamma$ due to a dynamical sign problem, which induces errors that increase exponentially with time. The classical mapping requires a large number of trajectories but the error is rather insensitive to the time needed for propagation.

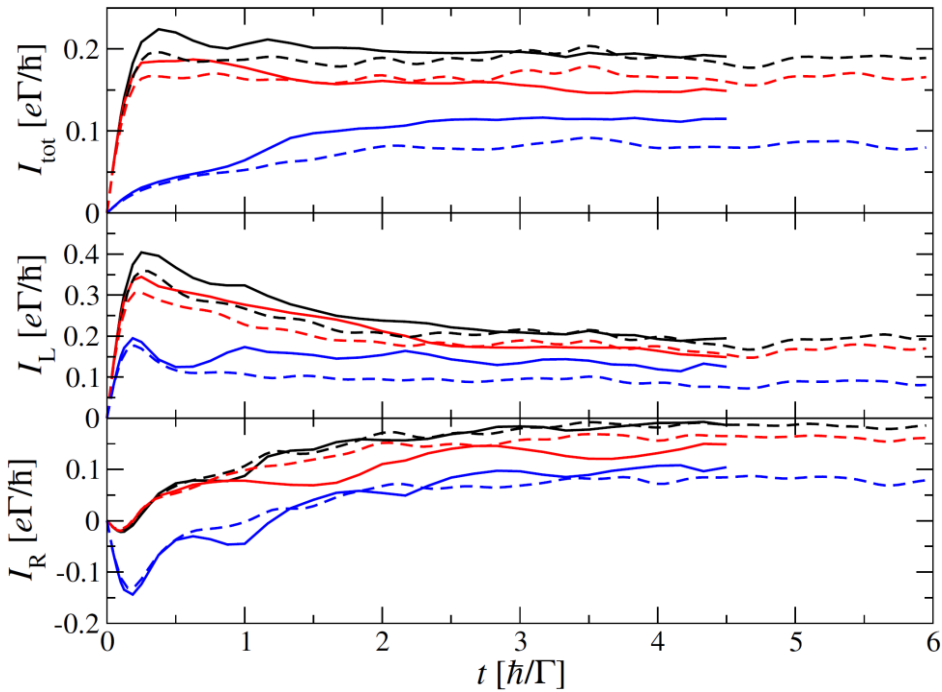

*Figure 2: Time dependent total (upper panel), left (middle panel) and right (lower panel) currents for the modified classical Hamiltonian of Eq. (18) (dashed curves) and RT-PIMC (solid curves). Black, red and green curves correspond to $\omega = 5, V_{sd} = 16$, $\omega = 3, V_{sd} = 16$ and $\omega = 5, V_{sd} = 2$, respectively. In all cases $\beta = 5, \gamma = 4$, and $\varepsilon_D = 0$. All the parameters are in units of $\Gamma$.*

The overall agreement between the classical mapping and the exact quantum results is more than satisfying. The mapping captures qualitatively most of the features observed in the time-dependent current. At short times ($t \approx \frac{2\pi}{\Gamma}$) it agrees well with the quantum results for low $V_{sd}$ while at longer times, as the currents levels off, the agreement is better for large $V_{sd}$. In some cases, we find that the coherent oscillations of the current with the vibrational frequency $\omega$, observed at intermediate times, are captured by the classical mapping, although the phase of the oscillations is sometimes incorrect. We note that rapid oscillations in the classical mapping results at frequencies higher than ω can be removed by further averaging.

Fig.3 shows the steady-state currents and the dot population for different vibrational frequencies and temperatures. At $k_B T = \frac{1}{5}$ (upper left panel of **Error! Reference source not found.**) the I-V characteristics exhibit a distinct step-like feature at $V_{s-d} = 2(\varepsilon_D + j\omega)$, where $j$ is an integer. This is clearly seen for both frequencies ($\omega = 3$ and $\omega = 5$). As discussed above, these step-like features, which are also present in the dot populations (right panel of **Error! Reference source not found.**), represent vibronic structures which are caused by the Franck-Condon effect. The comparison to the numerically exact RT-PIMC data shows that the classical mapping results are almost quantitative for the entire range of source-drain bias studied. Importantly, the mapping captures the step-like vibronic structures despite slightly over estimating their width. As the temperature increases to $k_B T = 1$, the step-like structure is mostly quenched (lower left panel of Fig. 3). This transition is also captured quantitatively by the classical mapping.

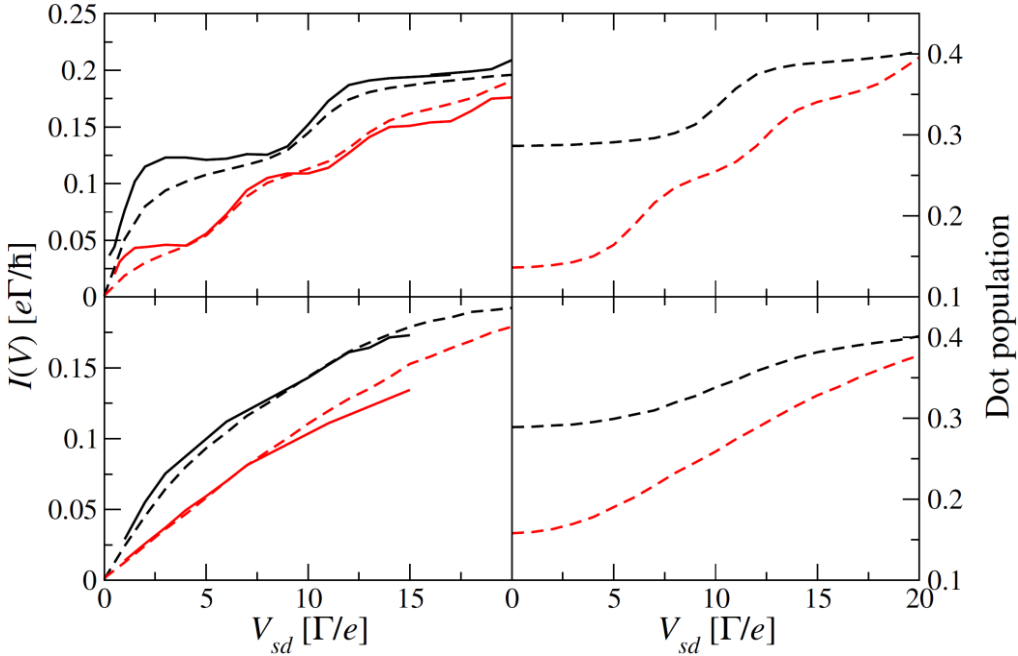

*Figure 3: Steady-state currents (left panels) and dot population (right panels) for $\omega = 5$ (black curves) and $\omega = 3$ (red curves). Shown are the exact RT-PIMC quantum results (solid lines) and the classical mapping results (dashed lines) for $\beta = 5$ (upper panels) and $\beta = 1$ (lower panels). In all cases $\gamma = 4$ and $\varepsilon_D = 0$. All the parameters are in units of $\Gamma$.*

It should be emphasized that the almost quantitative description of the vibronic steps in the I-V curves, which are related to the Franck-Condon effect, by a quasi-classical treatment is by no means a trivial result. As already mentioned above, these vibronic Franck-Condon structures are due to quantization of the vibrational DoF in the molecular resonance state and are missed in traditional quasi-classical treatments. This has been shown for transport in the Holstein model in **Error! Reference source not found.** and has also been investigated in detail for the related phenomenon of Franck-Condon structures in molecular absorption spectra.[88,89] The incorporation of these quantum effects in the present quasi-classical treatment is achieved by the special representation of the Hamiltonian (Eq. (22)) in combination with the binning procedure employed. While this special treatment is necessary within a quasi-classical implementation, it is expected that a fully semiclassical treatment (e.g. using a SC-IVR) of the Holstein model based on the Cartesian mapping, will capture these quantum effects directly, albeit at a significantly higher computational effort.

## IV.  CONCLUSIONS

In this paper, we have developed a classical Cartesian mapping approach suitable to describe vibrationally coupled charge transport in molecular junctions. As an example, we have considered the nonequilibrium Holstein model and studied the time-evolution of the current and dot occupation as the system relaxes to steady-state.

To capture the vibrational quantum effects within a quasi-classical treatment, we introduced a simple transformation rewriting the Holstein Hamiltonian in terms of electronic and vibrational occupation operators only, rather than separate creation and annihilation operators. Within this representation, quantization of the electronic and vibrational occupation was achieved by a multistage binning function, similar in spirit to the path taken for the Anderson impurity model. This proved crucial in describing the vibronic effects, in particular the phenomenon of Frank-Condon blockade (and Coulomb blockade in our previous study), giving rise to step-like features in the I-V characteristics. It also suggests a general framework to capture correlation effects in transport junctions using classical mappings.

The successful description of both Coulomb and Frank-Condon blockades within the Cartesian mapping approach relies mainly on quantizing the electron-electron/electron-vibrational coupling terms in the Hamiltonian. This, perhaps, is the Achilles heel of the approach, as it limits the class of models that can be studied. However, for suitable models, the approach has several appealing features, which may make it the "method of choice" for many applications in molecular transport junctions. Perhaps most notable is the linear scaling with system size and the potential application to complicated many-body Hamiltonians beyond the harmonic approximation. The limitation to a specific class of models may be avoided using a fully semiclassical treatment, which requires, however, a significantly higher computation effort.

**Acknowledgments**

This work was supported by the National Science Foundation Grant No. CHE-1148645 and by the Director, Office of Science, Office of Basic Energy Sciences, Chemical Sciences, Geosciences, and Biosciences Division, U.S. Department of Energy under Contract No. DE-AC02-05CH11231. EYW is grateful to The Center for Nanoscience and Nanotechnology at Tel Aviv University for a doctoral Fellowship. MT thanks the Chemistry Department at the University of California, Berkeley for a visiting Pitzer Professorship and W. H. Miller (UC Berkeley) and J. Neaton (Molecular Foundry, LBNL) for their hospitality. We also acknowledge a generou allocation of supercomputing time from the National Energy Research Scientific Computing Center (NERSC) and the use of the Lawrencium computational cluster resource provided by the IT Division at the Lawrence Berkeley National Laboratory.